\pdfoutput=1
\documentclass[a4paper,fleqn,usenatbib]{mnras}

\usepackage{newtxtext,newtxmath}

\usepackage[T1]{fontenc}
\usepackage{ae,aecompl}

\usepackage[dvipdfmx]{graphicx}	
\usepackage{amsmath}	
\usepackage{amssymb}	
\usepackage{booktabs} 

\newcommand{\unit}[1]{\ \mathrm{#1}}

\newcommand{\pa}[3][]{\frac{\partial^{#1} #2}{{\partial #3}^{#1}}}

\newcommand{\Msol}{\mathrm{M}_{\odot}}
\newcommand{\ord}[1]{\times 10^{#1}}

\newcommand{\average}[1]{\ensuremath{\langle#1\rangle} }

\newcommand{\enc}[1]{\left(#1\right)}

\usepackage{cancel}

\newcommand{\pink}[1]{\textcolor{black}{#1}}

\title[Magnetic Downflows in the Galactic Centre Region]{Magnetic Activity in the Galactic Centre Region -- Fast Downflows along Rising Magnetic Loops}

\author[Kakiuchi et al.]{
	Kensuke Kakiuchi$^{1,2}$\thanks{E-mail:kakiuchi@nagoya-u.jp}, 
	Takeru K. Suzuki$^{2,1}$, 
	Yasuo Fukui$^{1}$, 
	Kazufumi Torii$^{3}$, 
	\newauthor
	Rei Enokiya$^{1}$,
	Mami Machida$^{4}$, 
	and Ryoji Matsumoto$^{5}$
	\\
	$^1$Dept. of Physics, Nagoya University, Nagoya, Aichi 464-8692, Japan\\
	$^2$School of Arts \& Sciences, The University of Tokyo,3-8-1, Komaba, Meguro, Tokyo, 153-8902, Japan\\
	$^3$Nobeyama Radio Observatory, National Astronomical Observatory of Japan, 462-2, Nobeyama,\\
	Minamimaki, Minamisaku, Nagano, 384-1305, Japan\\
	$^4$Dept. of Physics, Faculty of Sciences, Kyushu University,744, Motooka, Nishi-ku, Fukuoka, 819-0395, Japan\\
	$^5$Dept. of Physics, Graduate School of Science, Chiba University, 1-33 Yayoi-cho, Inage-ku, Chiba 263-8522, Japan
}

\date{Accepted XXX. Received YYY; in original form ZZZ}

\pubyear{2018}

\begin{document}
\label{firstpage}
\pagerange{\pageref{firstpage}--\pageref{lastpage}}
\maketitle

\begin{abstract}
We studied roles of the magnetic field on the gas dynamics in the 
Galactic bulge by a three-dimensional global magnetohydrodynamical simulation data,
particularly focusing on vertical flows that are ubiquitously excited by magnetic activity.
In local regions where the magnetic field is stronger,
\pink{
it is frequently seen that fast downflows slide along inclined magnetic field lines that are associated with buoyantly rising magnetic loops. The vertical velocity of these downflows reaches $\sim 100$ km s$^{-1}$ near the footpoint of the loops by the gravitational acceleration toward the Galactic plane.
}
The two footpoints of rising magnetic loops are generally located at different radial locations
and the field lines are deformed by the differential rotation.
The angular momentum is transported along the field lines,
and the radial force balance breaks down.
As a result, a fast downflow is often observed only at the one footpoint
located at  the inner radial position. 
The fast downflow compresses the gas to form a dense region near the footpoint,
which will be important in star formation afterward. 
Furthermore,
\pink{
the horizontal components of the velocity are also fast near the footpoint because the downflow is accelerated along the magnetic sliding slope.
As a result, the high-velocity flow creates various characteristic features in a simulated position-velocity diagram, depending on the viewing angle. 
}
%
\end{abstract}




\section{Introduction}
%
The Galactic Centre (GC after here) is still an enigmatic region of the Galaxy, which includes a dense stellar system as well as the massive interstellar medium (ISM) consisting of molecular gas.
The inner part of the ISM is confined to Galactic longitude $l=\pm 2$ degrees and is observed as the Central Molecular Zone \citep[CMZ;][]{Morris1996} that is elongated along the Galactic plane. The CMZ exhibits high velocity widths of more than 200 km s$^{-1}$ and is characterized by a velocity pattern called ``parallelogram'' in the Galactic longitude-velocity ($l-v$) diagram in the molecular emission including the CO emission \citep{Bally1987}.
Recent observations report detailed structure of molecular clouds with complex streams \citep{Molinari2011,Kruijssen2015,Henshaw2016}.
The above mentioned large velocity dispersion of $\gtrsim 200$ km s$^{-1}$ indicates that
the CMZ has significant radial motion that is not described by rotation alone
\citep{Sawada2004}.
Such significant radial motion is unique to the GC and is not seen in the disc which shows only Galactic rotation of $\sim$ 220 km s$^{-1}$ with a small expansion of $\sim$ a few tens km s$^{-1}$ locally driven by super shells and possibly by HII regions.

Roles of a central stellar bar \citep{deVaucouleurs1978} have been widely discussed
in driving the radial motion
\citep{Binney1991,Morris1996,Koda2002,Baba2010}.
The basic mechanism that drives the radial motion is attributed to
two closed elliptical orbits called X1 and X2 orbits.
The bisymmetric gravitational potential excites non-circular streams that explain the parallelogram in the $l-v$ diagram \citep{Binney1991}.
The stellar bar that spans $l=\pm$ 20 degrees is well verified
by near infrared observations \citep{Matsumoto1982,Nakada1991,Weinberg1992,Whitelock1992,Dwek1995}
and the bar potential was shown to nicely explain the 3 kpc arm \citep{Oort1958} that is distributed in a much larger scale than the CMZ.
However, we have no observational evidence for a smaller-scale bar with a size of the CMZ;
the heavy extinction toward the Galactic centre hampers to reveal
asymmetric stellar distribution, in spite of the frequent discussion
on the bar potential model of the CMZ in the literature
\citep{Huttemeister1998,Rodriguez-Fernandez2008,Molinari2011,Krumholz2017}.
Moreover, it is quite difficult to reproduce the observed asymmetric feature of the parallelogram and complex flow
structures discussed above only by the bar potential.
These show that other ingredients also affect the gas dynamics in the GC region.

More recent studies showed that the magnetic field plays an important role in gas motion, possibly including that in the CMZ.
The original idea dates back to \citet{Parker1966}, which presented
a basic framework of the formation of clouds by magnetic buoyancy called Parker instability. The instability is a magnetic version of the Rayleigh-Taylor instability which creates a loop by the magnetic flotation in the vertical direction from the Galactic plane.
The floated gas in the loop eventually falls down toward the plane by the gravity and forms two footpoints of the ISM at the both ends of the loop, which explains cloud formation in the Galactic plane.
This instability is expected to be significant where the magnetic field is strong. 
\citet{Parker1966} argued that the instability will be more dominant in
the GC where the stronger gravity amplifies the magnetic field.
A number of numerical simulations for the Parker instability have been
performed to study general gas properties \citep{Matsumoto1988,Chou1997,Machida2013,Rodrigues2015}.

Another type of magnetic instability is magneto-rotational instability
(MRI hereafter) in a differentially rotating system \citep{Velikhov1959,Chandrasekhar1961,Balbus1991}.
MRI is powerful instability that drives horizontal flows in a disc;
the Parker instability and MRI are expected to excite three-dimensional (3D) complex gas flows. 
\citet{Machida2009} for the first time demonstrated by MHD simulations that the magnetic activity creates loop structure and
radial gas motion in the GC disc of a 1 kpc scale. 
Recently, \citet{Suzuki2015} showed by a MHD simulation that the
magnetic field is amplified more than a few 100 $\mu$G in the central 200
pc of the GC and radial flows are excited by the magnetic activity in a stochastic manner.  
They also successfully reproduced an asymmetric parallelogram shape
in simulated $l-v$ diagrams.
These works show that the magnetic activity offers a possible and
reliable alternative to the bar potential.

The magnetic driving of gas motion was observationally confirmed as two molecular loops in $l=$ 354 deg to 358 deg by \citet{Fukui2006}. 
The loops are likely located at $\sim$ 700 pc from the centre and
the field strength is estimated to be $\sim$~100 $\mu$G from the velocity dispersion of 100 km s$^{-1}$ in the footpoins of the loops.
This was a first observational confirmation of the prediction by \citet{Parker1966} and was followed by detailed molecular observations and
their analyses \citep{Fujishita2009,Torii2010b,Kudo2010}. 
Recently, \citet{Crocker2010} gave a stringent lower bound $> 50 \mu$G for the magnetic field strength within the central 400 pc region
by combining radio observation and the upper limit of $\gamma$-ray observation by EGRET.
Thus, it is a natural idea that the field becomes as large as 1 mG in the inner most part of a 200 pc radius where the CMZ is distributed. 
Complex structures, such as non-thermal filaments, which are likely
because of magnetic effect, are observed in the radio wavelength
range
\citep{Tsuboi1986, Oka2001}. 
If these features are formed by magnetic field, the expected field strength is 0.1 - 1 mG,
which is estimated from the equipartition between the magnetic pressure and the gas pressure
\citep{Yusef-Zadeh1984, Morris1989}.
Furthermore, \citet{Pillai2015} estimated the magnetic field strength of an infrared dark cloud G0.253+0.016,
which is known as ``the Brick'', is a few mG from the Chandrasekhar-Fermi method \citep{Chandrasekhar1953}.

With the background given above, our aim of the present paper is to
pursue detailed gas motion by using the simulation data of \citet{Suzuki2015}.
Our focus is particularly on the vertical motion of the gas. 
Vertical flows cannot be directly excited by the stellar bar
potential, and therefore they are unique to the magnetic activity.
The 3D-MHD simulation and its data are described in Section 2.
The results of our analysis are presented in Section 3.
We summarize the paper and discuss related issues in Section 4.

\section{The Data}
%

\begin{table}
	\centering
	\begin{tabular}{@{} llccc @{}} 
		\toprule
		&i&$M_i$ ($10^{10}\Msol$) & $a_i$ (kpc) & $b_i$ (kpc)\\
		\midrule\midrule
		SMBH&1& $4.4\times10^{-4}$& 0 &0\\ 
		Bulge&2& 2.05& 0 &0.495\\ 
		Disk&3& 25.47& 7.258 &0.52\\ 
		\bottomrule
	\end{tabular}
	\caption{Parameters adopted for \citet{Miyamoto1975} gravitational potential.}
	\label{table_grav}
\end{table}

\pink{We adopt the numerical dataset in the GC region} from a
3D-MHD simulation by \citet{Suzuki2015}.
This simulation was performed by solving ideal MHD equations
under \pink{an axisymmetric potential as external field;
the Galactic} gravitational sources consist of three components:
the supermassive black hole (SMBH) whose mass is $4.4\times10^6\Msol$ \citep{Genzel2010} at the GC (component $i=1$),
a stellar bulge ($i=2$) and a stellar disc ($i=3$).
\pink{The $i=2$ and 3 components are adopted from \citet{Miyamoto1975}, and we obtain}
%
\begin{align}
\Phi(R,z)=\sum_{i=1}^{3}{\frac{-GM_i}{\sqrt{R^2+\enc{a_i+\sqrt{b_i^2+z^2}}^2}}},\label{eqn_gpot}
\end{align}
where $R$ and $z$ are \pink{radius and height in the cylindrical coordinates}, respectively, and $M_i$, $a_i$ and $b_i$ are presented in Table \ref{table_grav}.

\citet{Suzuki2015} assumed locally isothermal gas, namely the temperature is spatially dependent but constant with time.
At each location, an equation of state for ideal gas, 
\begin{align}
p=\rho c_\mathrm{s}^2,
\end{align}
is satisfied, where $p$ is gas pressure, $\rho$ is density and
$c_\mathrm{s}\ (\propto {T}^{1/2})$ is ``effective'' sound speed. 
\citet{Suzuki2015} considered that $c_{\rm s}$ corresponds to the
velocity dispersion of clouds.
In the disc region ($R>$ 1.0 kpc),
\citet{Suzuki2015} adopted
from \citet[][]{Bovy2012},
\begin{align}
c_\mathrm{s,disc}=30 \unit{km\ s^{-1}}.
\end{align}
%
\pink{ In order to mimic
the observed large velocity dispersion  $\sim$ 100 km s$^{-1}$ within the bulge region of $R<$ 1.0 kpc
\citep{Kent1992} 
},
\citet{Suzuki2015} adopted 
\begin{align}
c_\mathrm{s,bulge}=0.6  v_\mathrm{rot,\star}=0.6\sqrt{R\pa{\Phi}{R}}\label{eqn_csbulge}
\end{align}
where $v_\mathrm{rot,\star}$ is the rotation speed of the stellar component.
\pink{
%
}
\pink{We would like readers to see \citet{Suzuki2015} for other detailed information of the numerical simulation.}

\section{Results}


\subsection{Vertical structures and flows}
\pink{In general, the magnetic filed tends to be more amplified}
in regions with stronger
differential rotation, because MRI and field-line stretching are more effective.
The differential rotation is partly strong
at the Galactic plane
in the bulge region, and the magnetic field
effectively amplified to 0.1-1 mG there as shown in \citet{Suzuki2015}. 
The stronger magnetic field is also expected to drive vertical flows by magnetic buoyancy
\citep{Parker1966} and the gradient of magneto-turbulence \citep{Suzuki2009,Suzuki2014}, which we focus on in this paper.

\begin{figure}
	\centering
	\includegraphics[clip,width=\hsize]{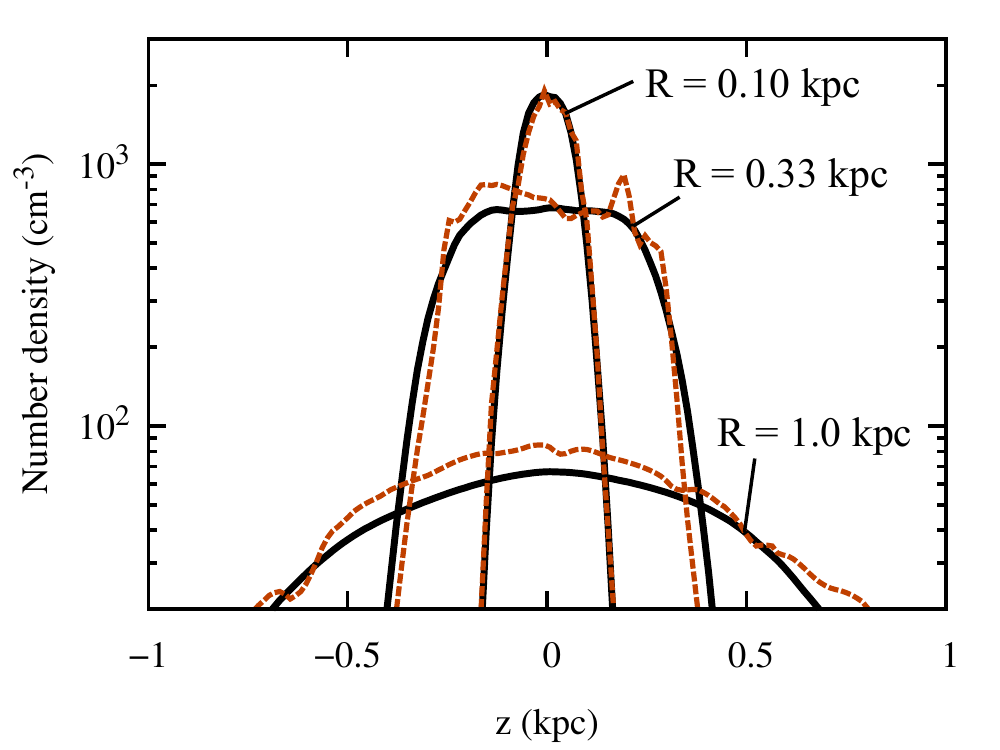}
	\caption{
		Vertical profile of the number density averaged over the azimuthal
		angle at different radial locations, $R=$0.10 kpc, 0.33 kpc, 1.00 kpc. 
		The solid lines denote the profile averaged over time from 399.5 to 402.5 Myr. 
		The dotted lines show snapshots at $t=400.10$Myr, at $\phi=0.0\unit{degree}$.
	}
	\label{fig_rho2pz}
\end{figure}

Figure \ref{fig_rho2pz} shows the vertical profile of the number
density at $R=0.10 \unit{kpc}$, 0.33 kpc, 1.00 kpc. 
The solid lines denote the profile averaged \pink{over azimuthal angle $\phi$ and} time from 399.5 to 402.5 Myr, which is the final stage of the simulation. 
The dotted lines show snapshots at $t=401.0$ Myr and $\phi= 0.0$ degree.
The structure of the averaged density indicates that the vertical thickness of the density distribution increases with $R$. 
The averaged profiles are almost symmetric with respect to the Galactic plane. 
On the other hand,
\pink{
	the \textit{non-averaged} snapshot profiles at $\phi=0$ degree (dotted lines)
	}
	show large asymmetric fluctuations,
	which implies that the distribution of the gas is perturbed locally by magnetic activity
	\pink{
		in a non-axisymmetric and temporal manner. 
	}

\begin{figure}
	\centering
	\includegraphics[clip,width=\hsize]{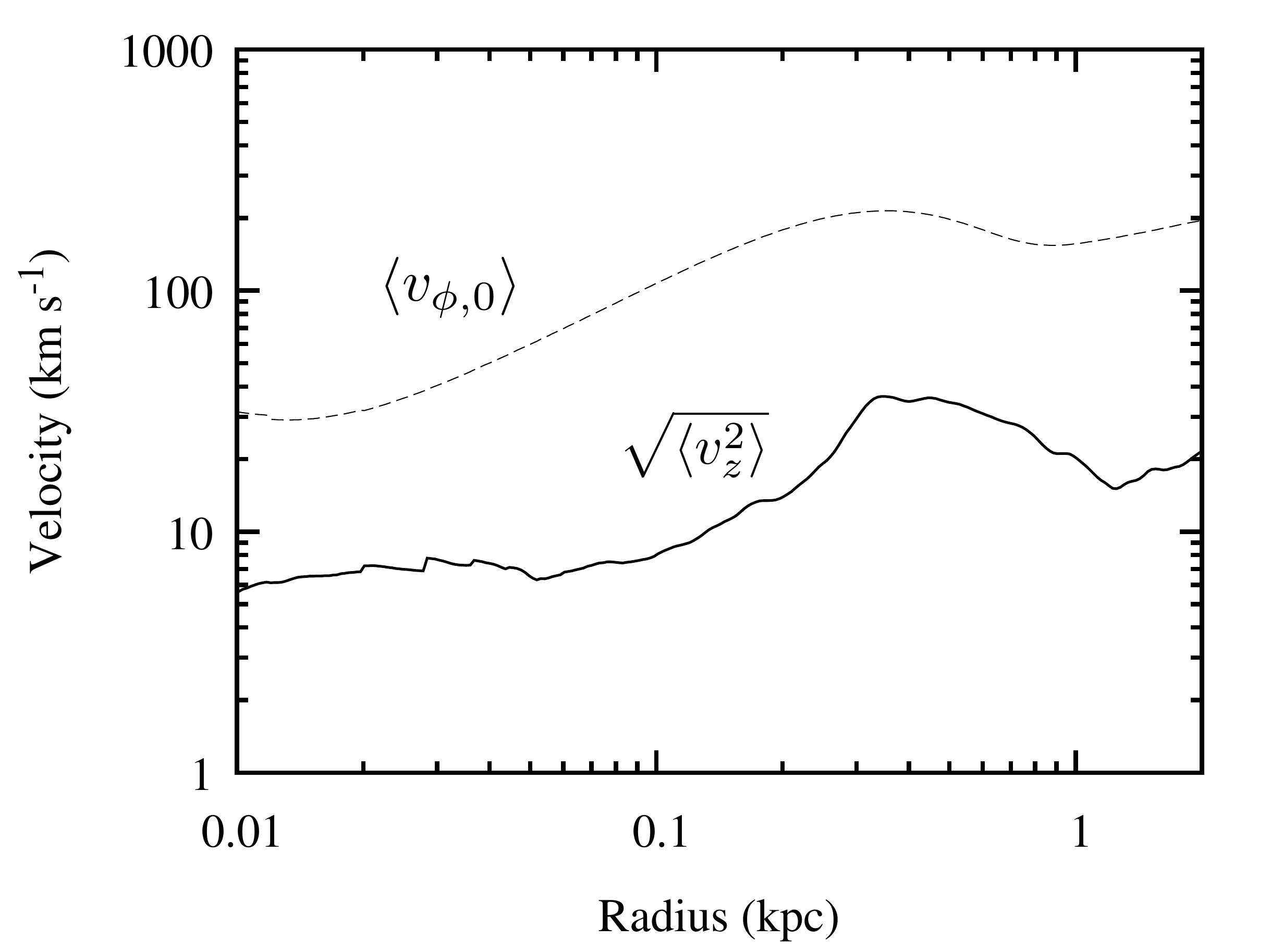}
	\caption{
		Radial distribution of root mean squared (rms) vertical velocity (solid line), 
		which is averaged over the azimuthal direction, 
		the vertical direction within $|z| < 1$ kpc, 
		and the time $t = 399.5-402.5$ Myr.
		The dashed line indicates the averaged azimuthal velocity at the initial condition. 
	}
	\label{fig_rdl2vz}
\end{figure}

Figure \ref{fig_rdl2vz} displays the radial distribution of root mean squared (rms) vertical velocity (solid line), 
which is averaged over azimuthal and vertical directions, 
and over the time $t = 399.5-402.5\unit{Myr}$.
Here, we take the average of a physical quantity, A, as follows:
\begin{align}
	\average{A}_{t,\phi,z}&=
	\frac{\int^{t_2}_{t_1}{dt} \int^{z_{2}}_{z_{1}}{dz} \int^{\pi}_{-\pi}{d\phi}\ A} 
	{2\pi (t_2-t_1)(z_2-z_1)}\label{eqn_average}
\end{align}
where we set $z_1=-1$ and  $z_2=+1$ kpc.
For the quantity concerning velocity, we  adopt the density-weighted average.
Following the equation \ref{eqn_average},
the average of rms velocity in vertical direction, $v_z$, is written as,
\begin{align}
	\sqrt{\average{v_z^2}}=\sqrt{\frac{\average{\rho v_z^2}}{\average{\rho}}}.
\end{align}
In Fig.\ref{fig_rdl2vz}, the dotted line indicates the averaged azimuthal velocity, 
	\begin{align}
		\average{v_{\phi,0}}={\frac{\average{\rho v_{\phi,0}}}{\average{\rho}}},
	\end{align}
at the initial condition.
This figure shows that the averaged vertical velocity has 10-20\%
speed of the initial rotational velocity.
This shows that the gas in the bulge region is mixed by the vertical motion.
The vertical velocity at the initial condition is zero,
because the distribution of the gas is determined
by the hydrostatic equilibrium. 
Thus,
we infer that these vertical flows are excited by the magnetic activity,
which we examine from now.

When we investigate roles of magnetic field, a plasma beta value,
\begin{align}
\beta=\frac{P_{\unit{gas}}}{P_{\unit{B}}}=\frac{8\pi\rho c_{\unit{s}}^2}{B^2}, \label{eqn_beta}
\end{align}
which is defined as the ratio of the gas pressure to the magnetic pressure, is a useful indicator.
In the low-$\beta$ plasma, the dynamics of gas is controlled by the
magnetic field. 
In this case,
\pink{the fluid motion driven by the magnetic activity is comparable to}
the Alfv\'en velocity,
\begin{align}
v_A &\sim 45\enc{\frac{B}{500 \unit{\mu G}}}\enc{\frac{n}{500 \unit{cm^{-3}}}}^{-1/2} \unit{km\ s^{-1}}\label{eqn_alfven},\\
&= 45\enc{\frac{B}{500 \unit{\mu G}}}\enc{\frac{\rho}{1.0\times10^{-21}\unit{g\ cm^{-3}}}}^{-1/2} \unit{km\ s^{-1}}\label{eqn_alfven2}.
\end{align}
where we adopted mean molecular weight, $\mu=1.2$, for the conversion between particle number density,
$n$, and mass density, $\rho$.
\pink{
	This estimate indicates that the typical velocity of flows driven by the magnetic activity is a few to several $10 \unit{km\ s^{-1}}$,
	which corresponds to the values obtained from Figure \ref{fig_rdl2vz}.
	If we compare equations (\ref{eqn_alfven}) and (\ref{eqn_alfven2}) to the sound velocity $\approx 100$ km s$^{-1}$ adopted in the simulation (equation \ref{eqn_csbulge}),
	the \textit{average} flow speed driven by magnetic activity is subsonic. However, we would like to note that flows could be supersonic in local regions with low density and strong magnetic field, which we actually observed in our simulation. 
}


\begin{figure}
	\centering
	\includegraphics[clip,width=\hsize]{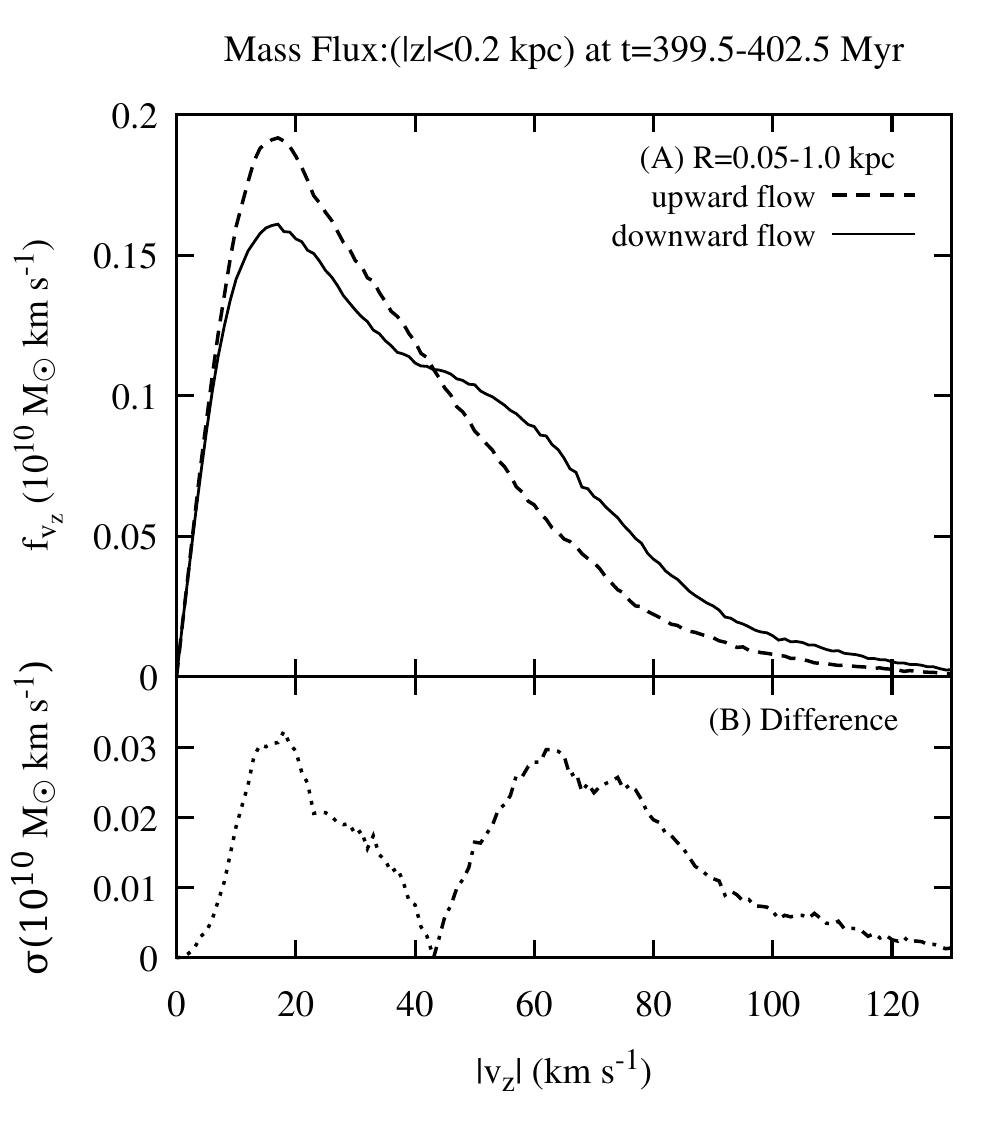}
	\caption{
		(A) The vertical velocity spectrum of mass flux in the bulge region ($R<1.0\unit{kpc}$, $|z|<0.2 \unit{kpc}$).
		The data are averaged over the time $t=399.5-402.5\unit{Myr}$.
		The horizontal axis denotes vertical velocity. 
		The vertical axis denotes vertical mass flux per $v_z$ bin.
		The solid (dashed) line indicates downward (upward) flows.
		(B)
		The difference between the solid and dashed lines in (A).
		The dotted (dot-dashed) line corresponds to the range in which upward (downward) flows dominate.
			}
	\label{fig_histgram}
\end{figure}

In order to study the vertical flows in a statistical sense, we separate the vertical motion of all the mesh points in
$R=0.05-1.0$ kpc and $|z|<0.2$ kpc into upflows
($v_z>0$ in $z>0$ or $v_z<0$ in $z<0$) and downflows
($v_z>0$ in $z<0$ or $v_z<0$ in $z>0$).
The top panel of Figure \ref{fig_histgram} shows the vertical velocity spectrum of mass flux,
\begin{align}
f_{v_z} =v_z \sum^{N}_{k=1} m_k,
\end{align}
where
\begin{align} 
m_k=\iint{\rho_k \textrm{d}z_k \textrm{d}S_k},
\end{align}
is the mass in each cell with number $k$. 
The data are averaged over the time $t=399.5-402.5\unit{Myr}$.
In this spectrum, we set the velocity bin $= 1 \unit{km\ s^{-1}}$.
The solid line indicates mass flux of the downflows and
the dashed line indicates that of the upflows. 

The bottom panel of Figure \ref{fig_histgram} presents 
the difference between the up-going mass flux and
down-falling mass flux,
\begin{align}
\sigma=|(f_{v_z})_u-(f_{v_z})_d |.
\end{align}
The dotted (dot-dashed) line corresponds to the velocity range
in which upward (downward) flows dominate downward (upward) flow. 
It is clearly seen that downward (upward) flows dominate
upward (downward) flows in the high (low) $v_z$ range;
the gas is lifted up slow and falls down fast.

We interpret this slow rise and fast drop from the magnetic activity and the gravity by the Galaxy potential.
We can estimate the velocity, $v_\textrm{gp}$, attained by the freefall from $z=0.25$ kpc to 0.05 kpc
at $R=0.4$ kpc, in the Galaxy potential, equation \ref{eqn_gpot}:
\begin{align}
v_\textrm{gp}(R,z)&= \sqrt{2 [\Phi(R',z')-\Phi(R,z)]}\label{eqn_vpot1}\\[2mm]
&\sim 100 \unit{km\ s^{-1}} \label{eqn_vpot2}\\
; \unit{at}\ R,R'&=0.4 \unit{kpc},\ z=0.25 \rightarrow z'=0.05 \unit{kpc}.\notag
\end{align}
This freefall velocity is moderately faster than the Alfv\'en velocity estimated in equation \ref{eqn_alfven}.
Therefore, it is consistent with the interpretation that
the gas is lifted up by the magnetic activity such as
Parker instability \citep{Parker1966,Parker1967a} and falls down by the gravity.

\begin{figure}
	\centering
	\includegraphics[clip,width=\hsize]{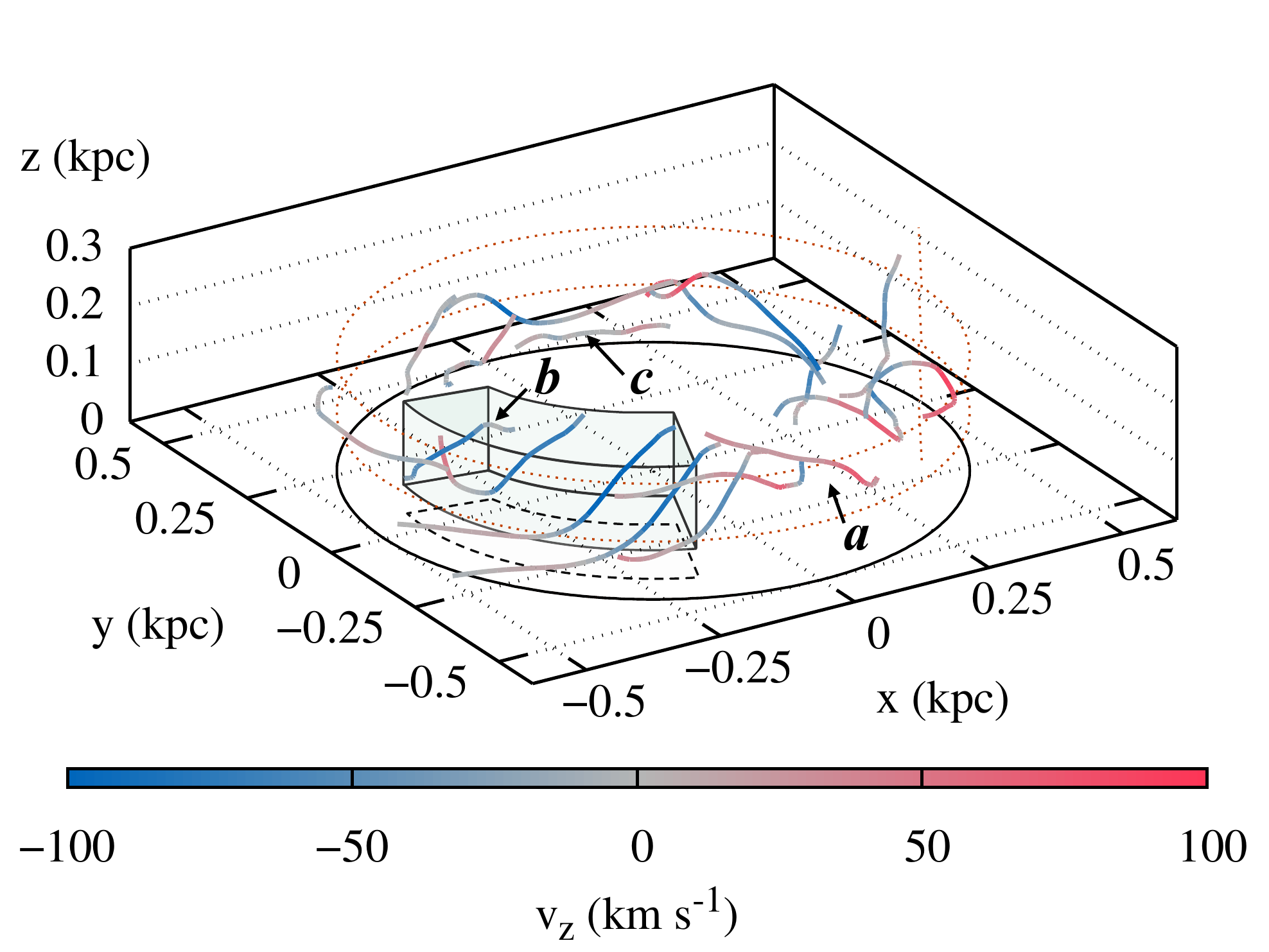}
	\caption{
		Overview of trajectories of various fluid elements in $R=0.3-0.6 \unit{kpc}$.
		Each line indicates the track of the fluid parcels from $t =399.5\unit{Myr}$ to $404.5\unit{Myr}$. 
		\pink{Note that those lines are different from the configuration of magnetic field lines.}
		Colors denote vertical velocity; redder colors correspond to upward flows and bluer colors correspond to downward flows.
		This figure clearly shows ubiquitous vertical flows.
		The \pink{boxed} region shows the Region X.
	}
	\label{fig_traject}
\end{figure}
\begin{figure}
	\centering
	\includegraphics[clip,width=\hsize]{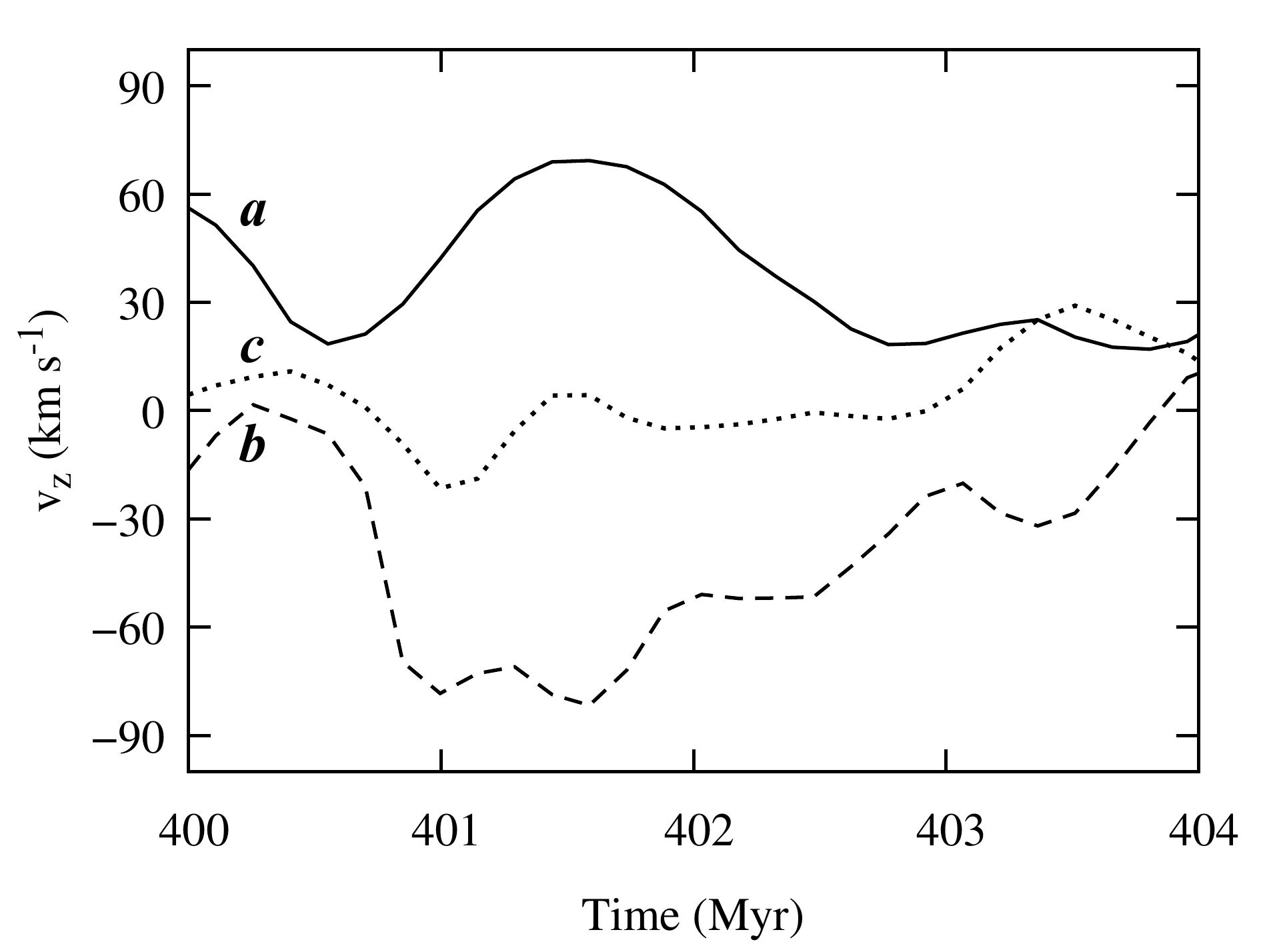}	
	\caption{
		Time evolution of vertical velocity of different fluid elements.
		\pink{
			Labels 'a','b' and 'c' correspond to the trajectories shown in Figure \ref{fig_traject}.
		}
		\pink{
			Since we focus on flows in the upper hemisphere, positive (negative) $v_z$ corresponds to upward (downward) flows.
		}
		}
	\label{fig_traject2}
\end{figure}

Figure \ref{fig_traject} presents
the trajectories of multiple fluid elements
in $R = 0.3 - 0.6 \unit{kpc}$.
Each line indicates the track of the fluid elements from $ t = 399.5 \unit{Myr}$ to $404.5 \unit{Myr}$;
\pink{They are different from the configuration of the snapshot of magnetic field lines, which will be shown in Sections 3.2 \& 4.3.}
Colors denote vertical velocity; redder colors correspond to upward flows and bluer colors correspond to downward flows.
This figure clearly shows ubiquitous vertical flows.

The fluid elements rotate in the
clockwise direction approximately with $\approx 150-200 \unit{km\ s^{-1}}$,
which is determined by the radial force balance
between the inward gravity and the centrifugal force;
the trajectories follow roughly one third of the rotation during the period of 5.0 Myr,
which is clearly seen in Figure \ref{fig_traject}.
Figure \ref{fig_traject} further shows that the fluid elements show
radial and vertical motion in addition to the background rotation;
some fluid elements show quite large vertical displacements of
$0.05-0.2$ kpc.

We selected three trajectories (a, b, c) from Figure \ref{fig_traject} and
show the time evolution of the vertical velocity
of these fluid elements in Figure \ref{fig_traject2}.
This shows that the velocity of the upgoing fluid element
(a; solid line) is about $20-60 \unit{km\ s^{-1}}$ 
and the velocity of the falling fluid element
(b; dashed line) is about $20-80 \unit{km\ s^{-1}}$.
Since the rotational velocity at $R=0.3-0.6 \unit{kpc}$ is $150-200 \unit{km\ s^{-1}}$, 
the vertical velocity reaches $\sim$ the half of the rotation velocity.
Apart from the monotonic upward (solid) and downward (dashed) flows,
the fluid element indicated by the dotted line (c) shows a fluctuating behavior around $v_z=0$.
\pink{
	The magnetic energy, $B^2/4\pi$, near the fluid element,
	``c'', is about an order of magnitude smaller than
	the magnetic energy near the fluid elements, ``a'' and ``b''.
	This implies that the vertical flows are closely related to the magnetic activity.
}
Although these are not shown in Figures \ref{fig_traject} \& \ref{fig_traject2},
some fluid parcels fall with nearly the freefall speed and cross the Galactic plane.

In the following section,
we focus on very fast downflows
(see the boxed region in Figure \ref{fig_traject})
and investigate how magnetic activity plays a role in driving vertical flows.

\begin{figure*}
	\centering
	\includegraphics[width=0.48\hsize]{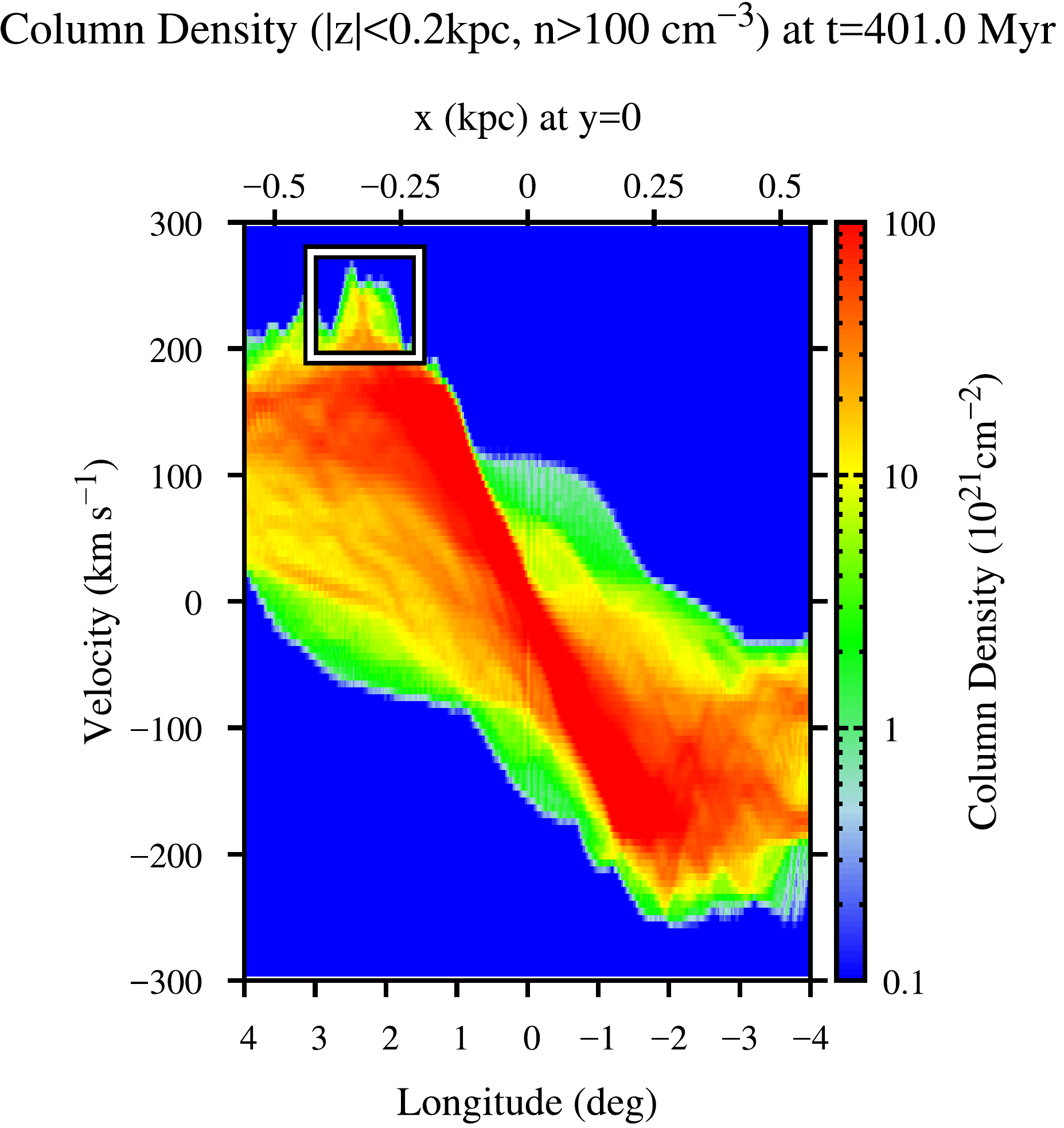}
	\hspace{0.02\hsize}
	\includegraphics[width=0.48\hsize]{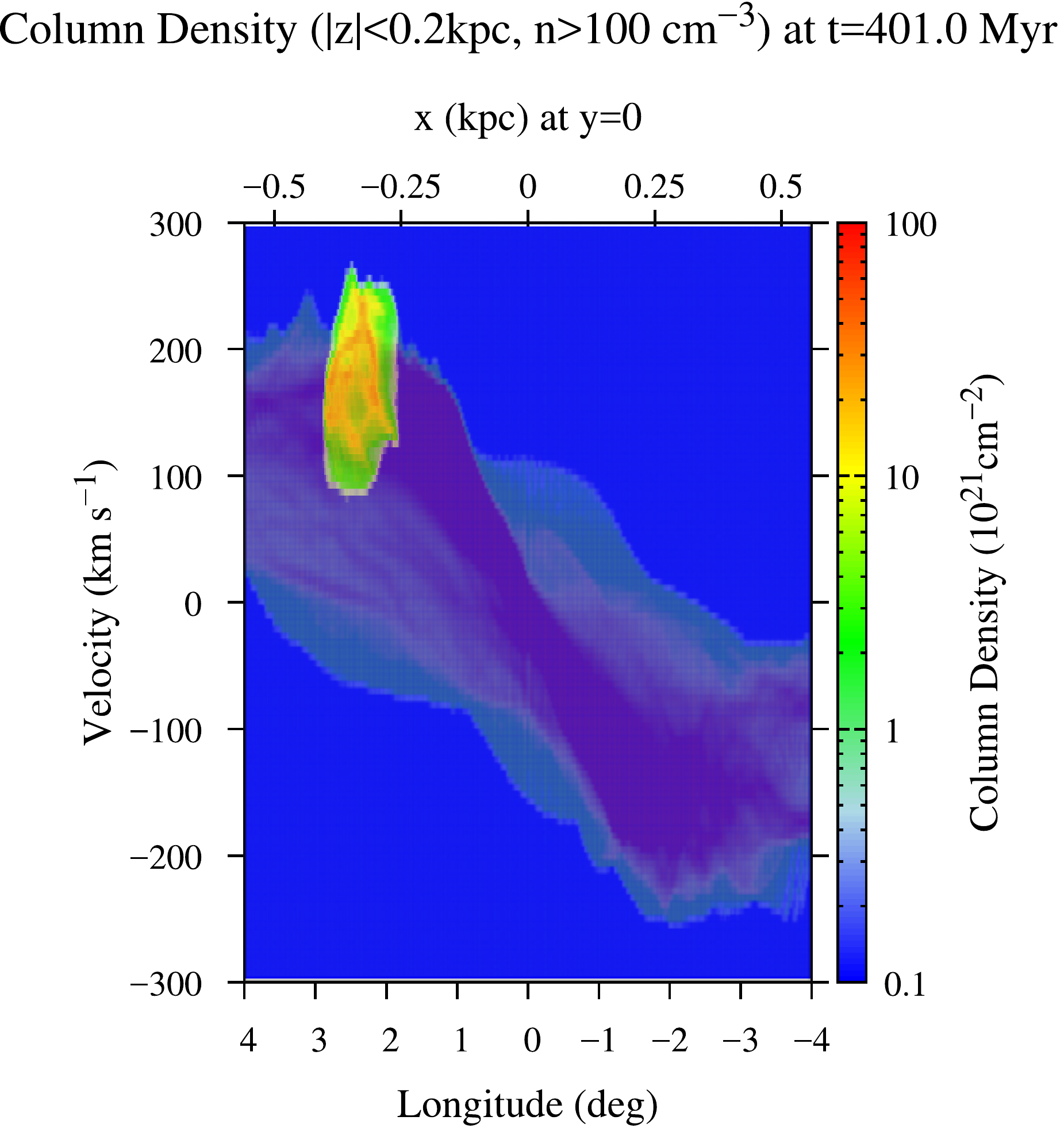}
	\caption{
		(Left) Galactic longitude--Line of sight velocity ($l-v$) diagram from observed LSR (Local Standard of Rest) at $t=401.0 \unit{Myr}$.
		The color contour denotes column density in units of $10^{21}\unit{cm^{-2}}$,
		where we use the grid points with high-density regions, $n> 100 \unit{cm^{-3}}$, in $|z| <0.2 \unit{kpc}$ to integrate along the direction of line of sight. 
		(Right) The area constructed from Region X is highlighted.
	}
	\label{fig_lv_diagram}
\end{figure*}

\subsection{Rising loop, Downflow, Compression}
In this subsection, we examine a local region that shows rapid downflows 
in $\phi=165-230^{\circ}$($\phi=0^{\circ}$ corresponds to the $+x$ axis),
$R=0.3-0.4$ and $z=0.05-0.2$ kpc (Figure \ref{fig_traject}),
which we call ``Region X''. 
These downflows start at $z\approx 0.2$ kpc and reach near the Galactic plane after 2 -- 5 Myr,
which corresponds to 1/4--1/2 of the rotation
period at $R\approx 0.3$ kpc. 

The left panel of Figure \ref{fig_lv_diagram} shows a $l-v$ diagram
observed from the LSR (Local Standard of Rest) at $t=401.0$ Myr.
\pink{Here we assume that the solor system rotates}
with 240 $\unit{km\ s^{-1}}$ in the clockwise direction
at $R=8.0 \unit{kpc}$ \citep{Honma2012}, $z=0.0 \unit{kpc}$ and $\phi=270^{\circ}$
(to the - y direction).
The color contour denotes column density in units of $10^{21}\unit{cm^{-2}}$
integrated along the direction of line of sight,
whereas the grid points that satisfy $n>100$ cm$^{-3}$ and $|z|<0.2$ kpc are adopted for the integration.  

Interestingly, the local rapid downflow region corresponds to the high $v$ peak in the white box of $l-v$ diagram in the left panel of Figure \ref{fig_lv_diagram}.
In the right panel, we highlighted the area constructed only from
Region X, which shows large velocity dispersion $> 150$ km s$^{-1}$
owing to the downflows in this local region.

\begin{figure*}
	\centering
	\includegraphics[clip,width=0.48\hsize]{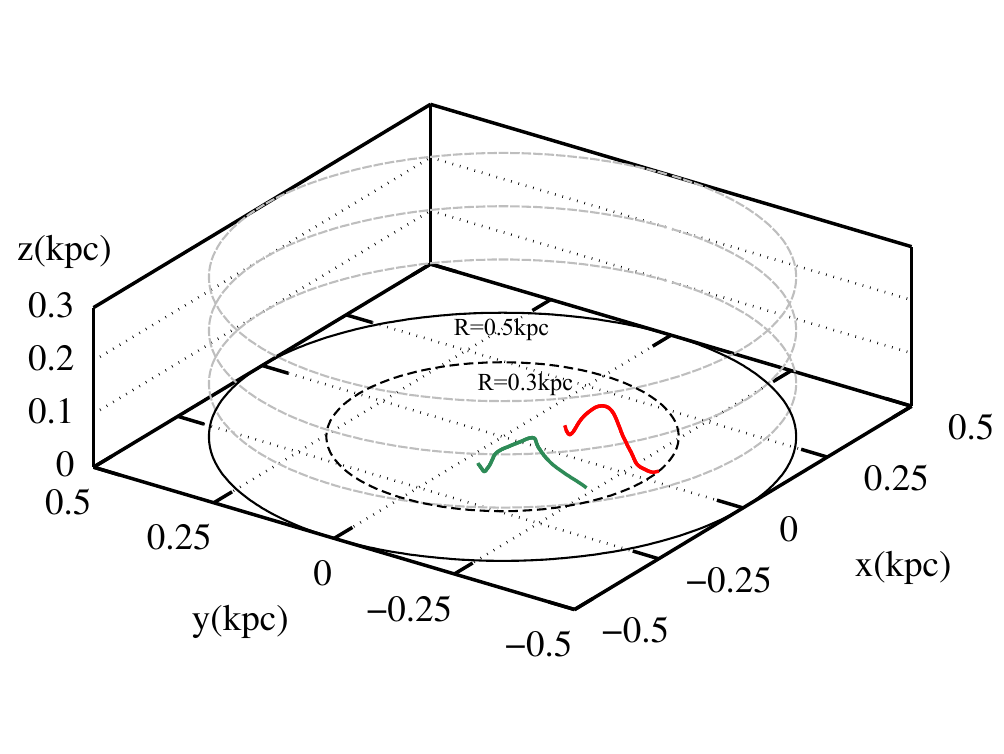}
	\includegraphics[clip,width=0.48\hsize]{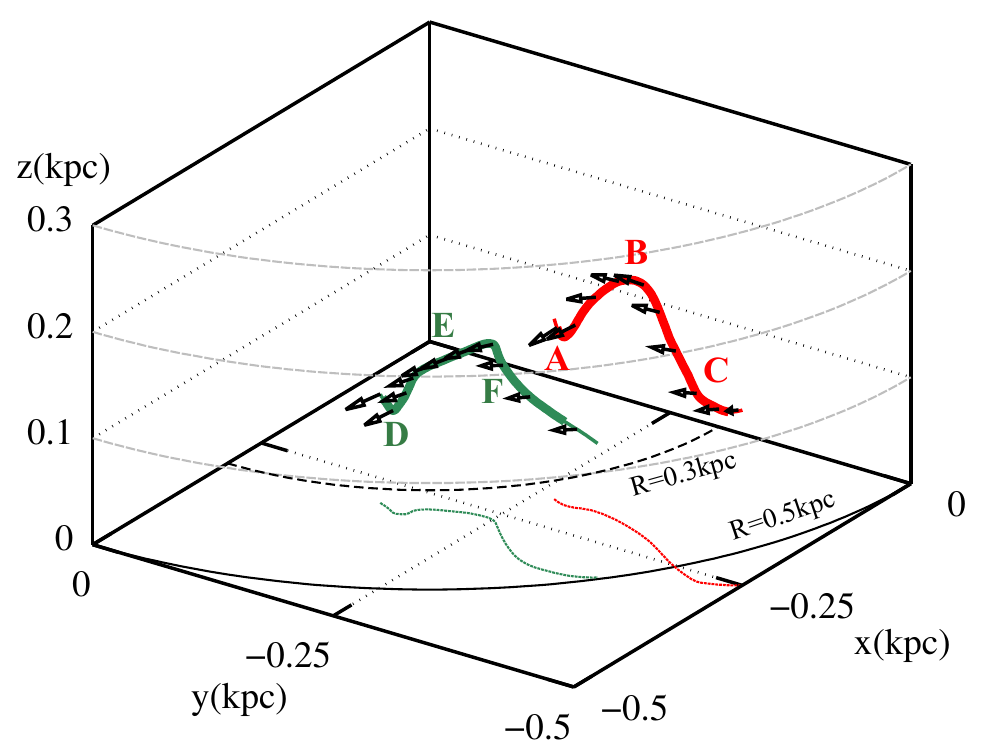}
	\caption{
		MF1 (red) and MF2 (green) at $t=401.0$ Myr.
		The right panel zooms in the left panel.
		Thin lines denote projected lines on x-y plane of MF1(red) and MF2 (green).
	}
	\label{fig_magneto2}
\end{figure*}

\begin{figure}
	\centering
	\includegraphics[clip,width=0.90\hsize]{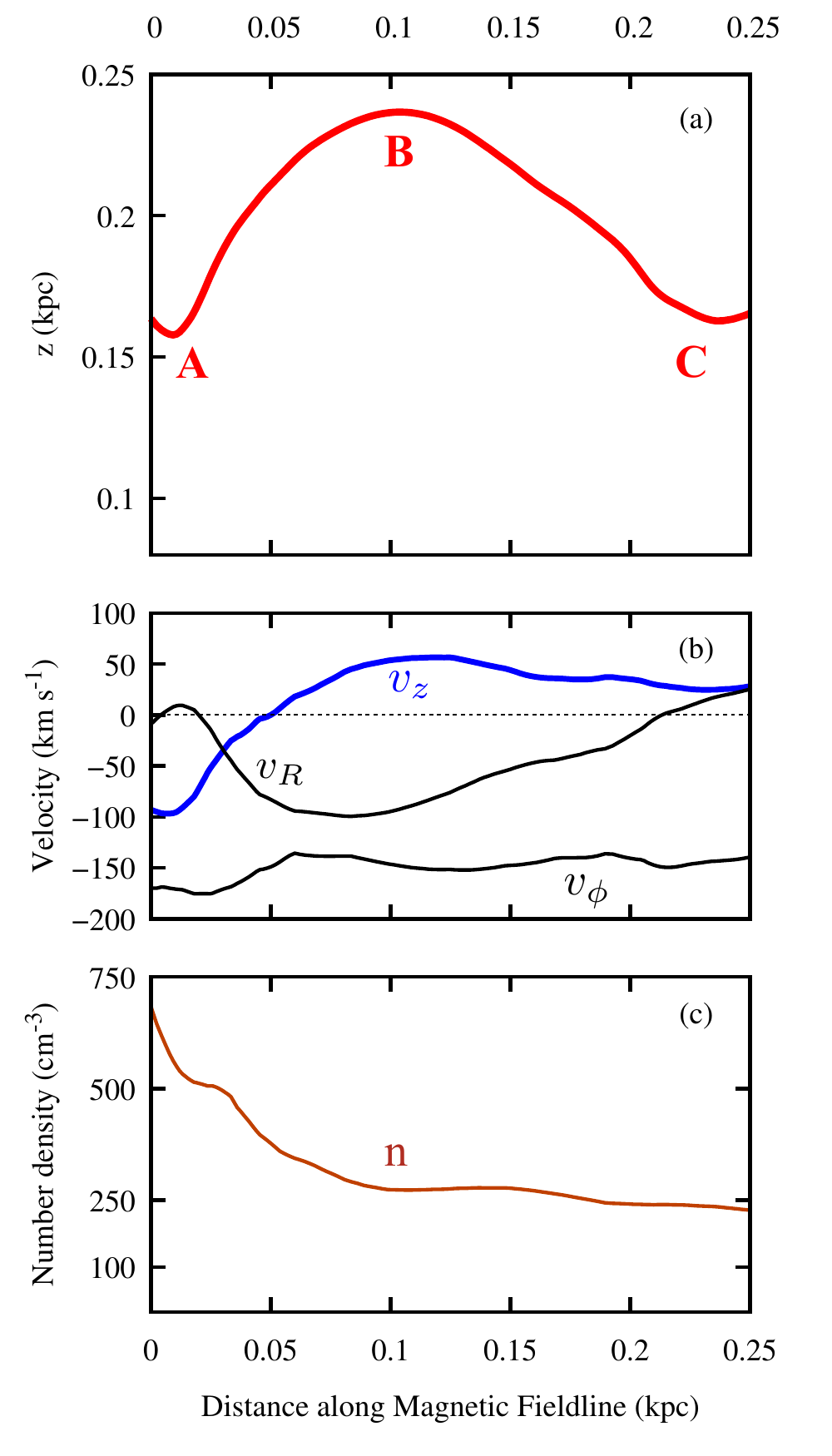}
	\caption{
	Physical quantities along the magnetic field line MF1. 
	(a) The height of the MF1 (solid line). 
	The alphabets (e.g., A, B, C) correspond to those in Figure \ref{fig_magneto2}.  
	(b) The velocity field at the each components. 
	The blue line denotes the vertical velocity. 
	The black lines denote the horizontal velocity, $v_{R}$ and $v_{\phi}$. 
	(c) The number density (solid).
	}
	\label{fig_loop}
\end{figure}
\begin{figure}
	\centering
	\includegraphics[clip,width=0.90\hsize]{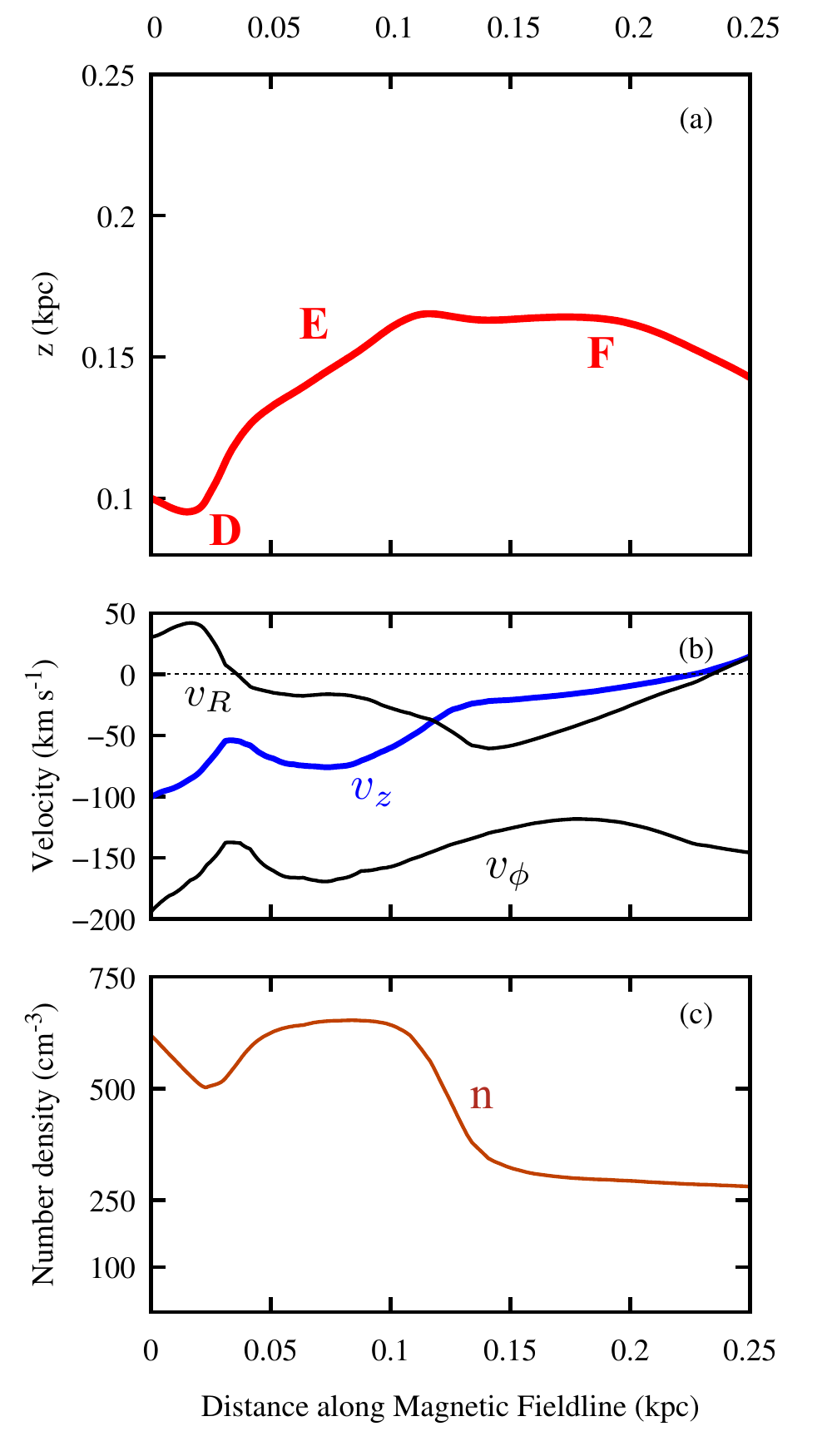}
	\caption{
		Same as Figure \ref{fig_loop} but for MF2. 
	}
	\label{fig_loop2}
\end{figure}

We show selected magnetic field lines in this Region X with rapid flows at
$t=401.0$ Myr in Figure \ref{fig_magneto2}. 
In the right panel, we picked up two magnetic field lines that exhibit loop-like configurations,
``Magnetic Field line 1 (MF1)''  shown by red and ''Magnetic Field line 2 (MF2)'' by green.
The two footpoints of MF1 and MF2 are anchored at different radial locations.

Figures \ref{fig_loop} and \ref{fig_loop2} present various physical quantities along MF1 and MF2, respectively. 
The top panel (a) of Figures \ref{fig_loop} and \ref{fig_loop2} shows the vertical height, $z$, (solid line).
The alphabets correspond to the locations shown in Figure \ref{fig_magneto2}.

The panel (b) of Figures \ref{fig_loop} and \ref{fig_loop2} presents the three components of the velocity. 
We would like to note that they are the snapshots at $t=401.0$ Myr,
which are different from the trajectory of passively moving fluid elements. 
Figure \ref{fig_loop}(b) shows that the loop-top region (``B'') of MF1 is upgoing with $\approx 50-60$ km s$^{-1}$.
Therefore,  this MF1 is still rising to a higher altitude. 
In the Region X, the averaged strength of magnetic field is
about $0.2-0.3 \unit{mG}$, and the Alfv\'{e}n velocity is $\approx 20$ km s$^{-1}$ (see also equation \ref{eqn_alfven}).
While the typical rising velocity by magnetic buoyancy is the Alfv\'{e}n velocity \citep{Parker1966,Parker1967a},
our result for MF1 shows that the former is a few times faster than the latter. 
Therefore, we expect that, in addition to the magnetic buoyancy, other mechanisms,
such as effective pressure of magnetic turbulence \citep{Suzuki2009},
should cooperatively work in lifting up the loop.
On the other hand, near the footpoint region, ``A'', of MF 1, the gas
falls down along the magnetic field with 50 -- 100 km s$^{-1}$, which
is comparable to the freefall velocity in the Galactic potential
(see equation \ref{eqn_vpot2}).  
Figure \ref{fig_loop2}(b) also shows that the gas falls down to
the Galactic plane with $v_z > 50$ km s$^{-1}$ near the footpoint
``D'' of MF2.

Another interesting feature is that the downward flow is seen only at the small $R$ side.
This is because the direction of the radial flow is inward ($v_{R}<0$) along MF1.
The footpoints of MF1 are anchored at different $R$ locations, which rotate at different angular velocities.
The inner footpoint, which rotates faster, is decelerated because the
magnetic connection suppresses the differential rotation (Figure
\ref{fig_loop}(b)) by the outward transport of angular momentum. This
is the same physical picture to the
MRI \citep{Balbus1991,Balbus1998}.
As a result, the radial force balance breaks down owing to the decrease of the centrifugal force.
The gas moves inward to the GC, which further accelerates the infall motion along the magnetic sliding slope.
If we follow this picture, the gas near the outer footpoint region moves outward. 
However, the overall rotation velocity of MF1 is smaller than the background rotation,
the positive $v_{R}$ at the outer footpoint is not as fast as the negative $v_{R}$ at the inner footpoint.
As a result, the gas slides up along the outer (right side in Figure \ref{fig_loop}) slope.

The strong downflow to the inner footpoint  near ``A'' is blocked by the counter stream from a neighbouring magnetic loop.
This downflow finally compresses the gas near the footpoint region.
Figure \ref{fig_loop}(c) actually shows that the density near the location ``A'' is 2-3 times higher than the density at the loop top (``B'').

\begin{figure}
	\centering
	\includegraphics[clip,width=\hsize]{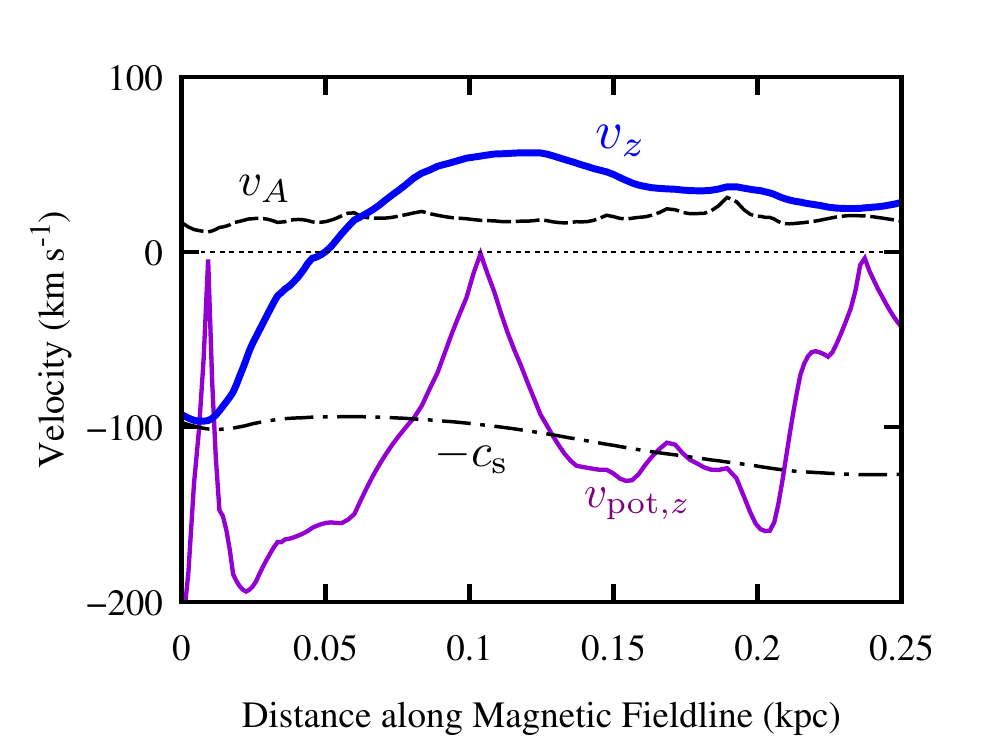}
	\includegraphics[clip,width=\hsize]{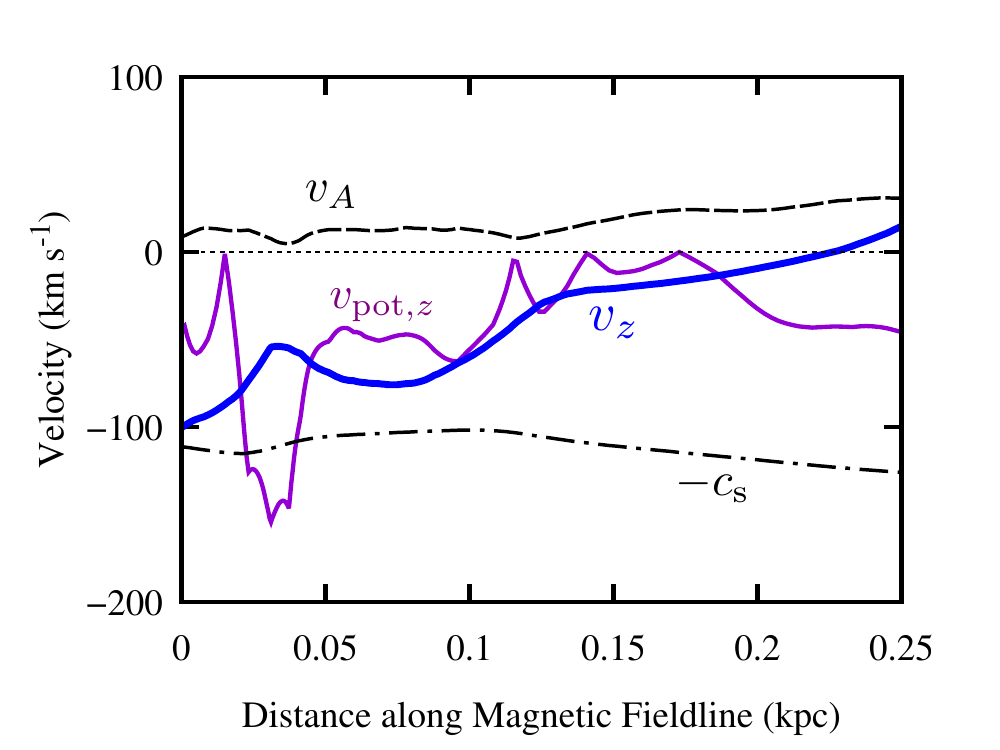}
	\caption{
	Comparison between the simulated $v_z$ (thick blue solid) and the kinematical estimate, $v_{{\rm pot},z}$  (thin purple solid; eq. \ref{eqn_vpotz}).
	The Alfven (dashed) and sound (dot-dashed) velocities are also shown.
	The top and bottom panels show MF1 and MF2, respectively. 
	}
	\label{fig_loop_cmpr}
\end{figure}

We would like to compare the velocity of downflows
along MF1 and MF2 to the velocity obtained by the gravity.
We can roughly estimate the velocity along an inclined magnetic field
line: 

\begin{align}
v_\textrm{pot,z} = -v_\textrm{gp}(R,z)\times|\sin{\Theta}|,
\label{eqn_vpotz}
\end{align}
where 
\begin{align}
	\sin{\Theta}=\frac{B_z}{|B|}=\frac{B_z}{\sqrt{B_R^2+B_\phi^2+B_z^2}},
\end{align}
and $\Theta$ is the inclination angle of the field line with respect
to the Galactic plane. 
We set the reference point of $v_{\rm gp}=0$ at $(R,z)=(0.64\;{\rm kpc},0.6\;{\rm kpc})$
for MF1 (Figure \ref{fig_loop}) and $(R,z)=(0.75\;{\rm kpc},0.32\;{\rm kpc})$ for MF2 (Figure 8).
The negative sign in equation \ref{eqn_vpotz} is for downward flows ($v_z < 0$).
The purple lines in Figure \ref{fig_loop_cmpr}
indicate $v_{{\rm pot},z}$ estimated in this manner.

Figure \ref{fig_loop_cmpr} shows that
the qualitative trend of $v_z$ is roughly followed by $v_{{\rm pot},z}$.
However, the detailed profile cannot be reproduced because the kinematical argument
based on equations \ref{eqn_vpot1} and \ref{eqn_vpotz} does not take into account the gas pressure and the Lorentz force. 
For instance, although $v_{{\rm pot},z}$ of MF1 predicts the acceleration of $\approx 200$ km s$^{-1}$ 
from the loop top, ``B'', to the footpoint, ``A'', 
this estimate is considerably larger than the actual downward acceleration of $v_z$, 
because MHD effects suppress the rapid downflow in reality.

2D MHD simulation \citep{Matsumoto1990} shows that downflows along a buoyantly rising loop excite shock waves near the footpoints.
Although the top panel of Figure \ref{fig_loop_cmpr} indicates that $v_z$ is comparable to the sound speed, $c_{\rm s}$, near the footpoint ``A'', we do not observe a typical shock structure, e.g. a jump of velocity and density. However, this may be because the adopted $c_{\rm s}$ is too large in order to take into account random velocity of clouds in addition to the physical sound speed \citep{Suzuki2015}.

\begin{figure}
	\centering
	\includegraphics[clip,width=\hsize]{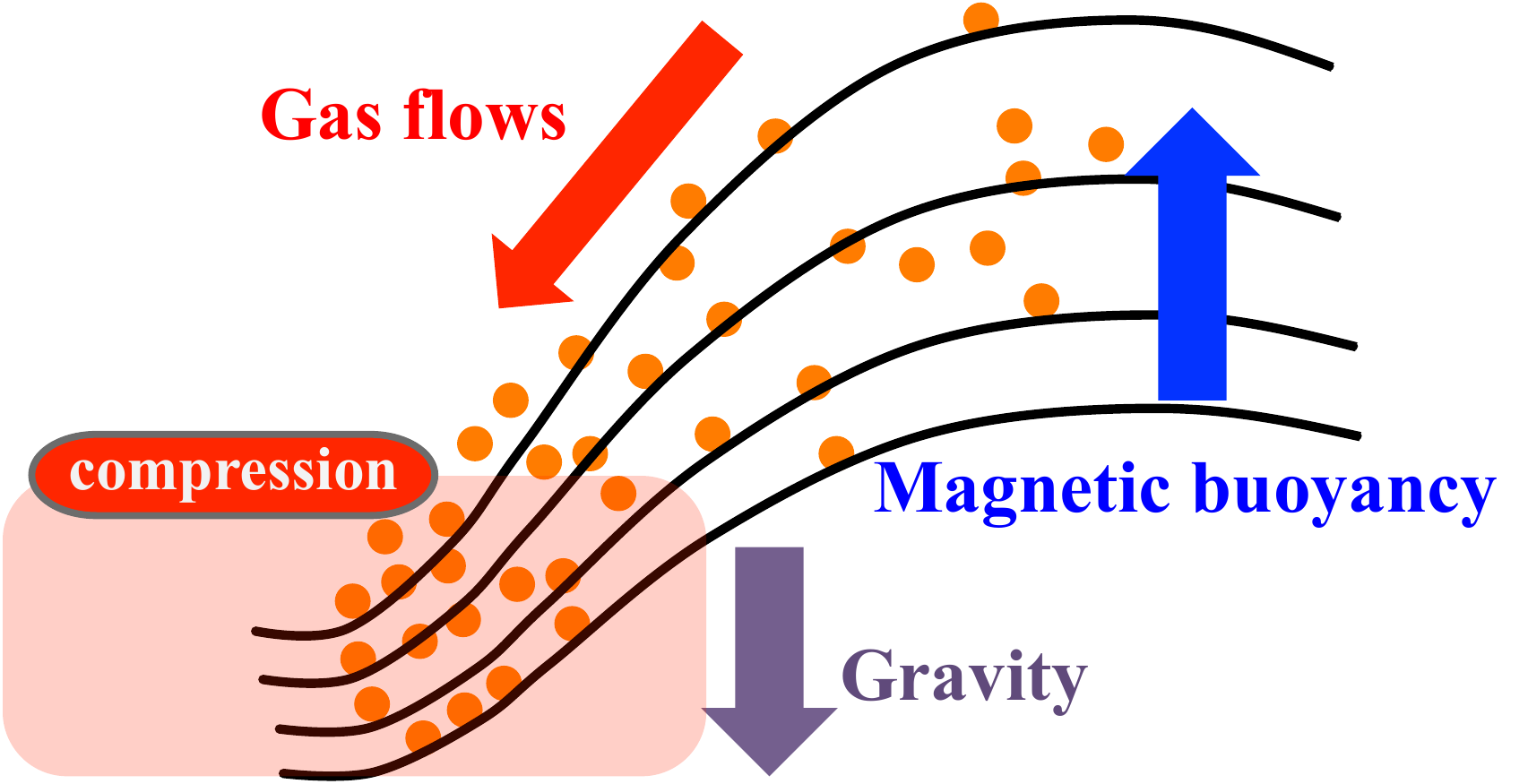}
	\caption{
	A conceptual image of a rising loop and the associated magnetic sliding slope. 
	}
	\label{fig_slope}
\end{figure}

Figure \ref{fig_slope} shows a conceptual view that summarized the results in this subsection. 
The magnetic field lines
are lifted up by the magnetic buoyancy (Parker instability) and
form a loop-like structure.  
The gas falls down along the inclined magnetic field, ``a magnetic sliding slope'',
and is finally compressed near the footpoint. 
We expect that the compressed gas will be eventually a seed of a molecular cloud \citep{Torii2010b},
although our simulation cannot treat such low-temperature gas and
cannot follow the formation of molecular clouds.

\section{Discussion}
\subsection{Dependence on viewing angle}
\begin{figure*}
	\centering
	\includegraphics[width=0.48\hsize]{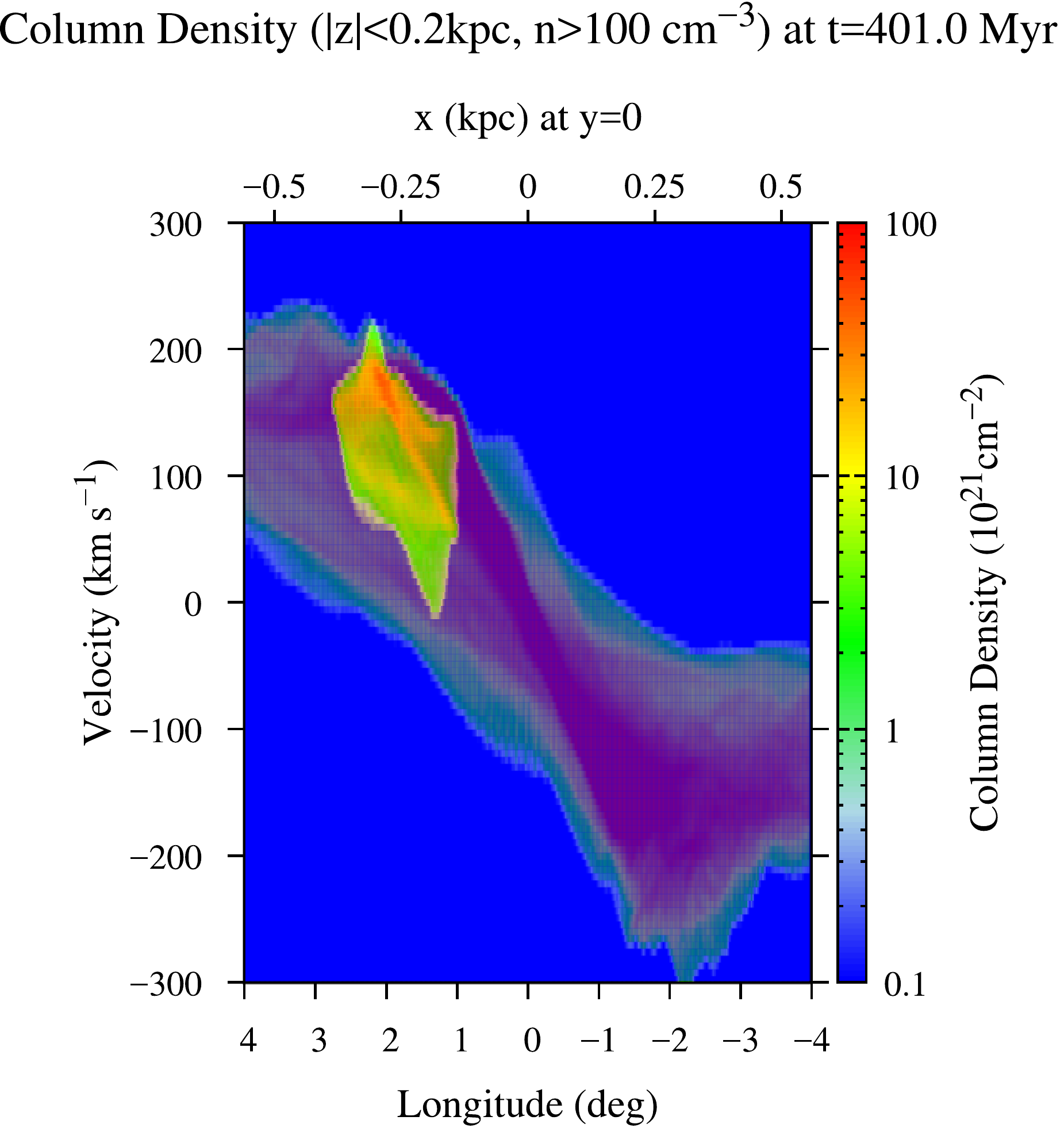}
	\hspace{0.02\hsize}
	\includegraphics[width=0.48\hsize]{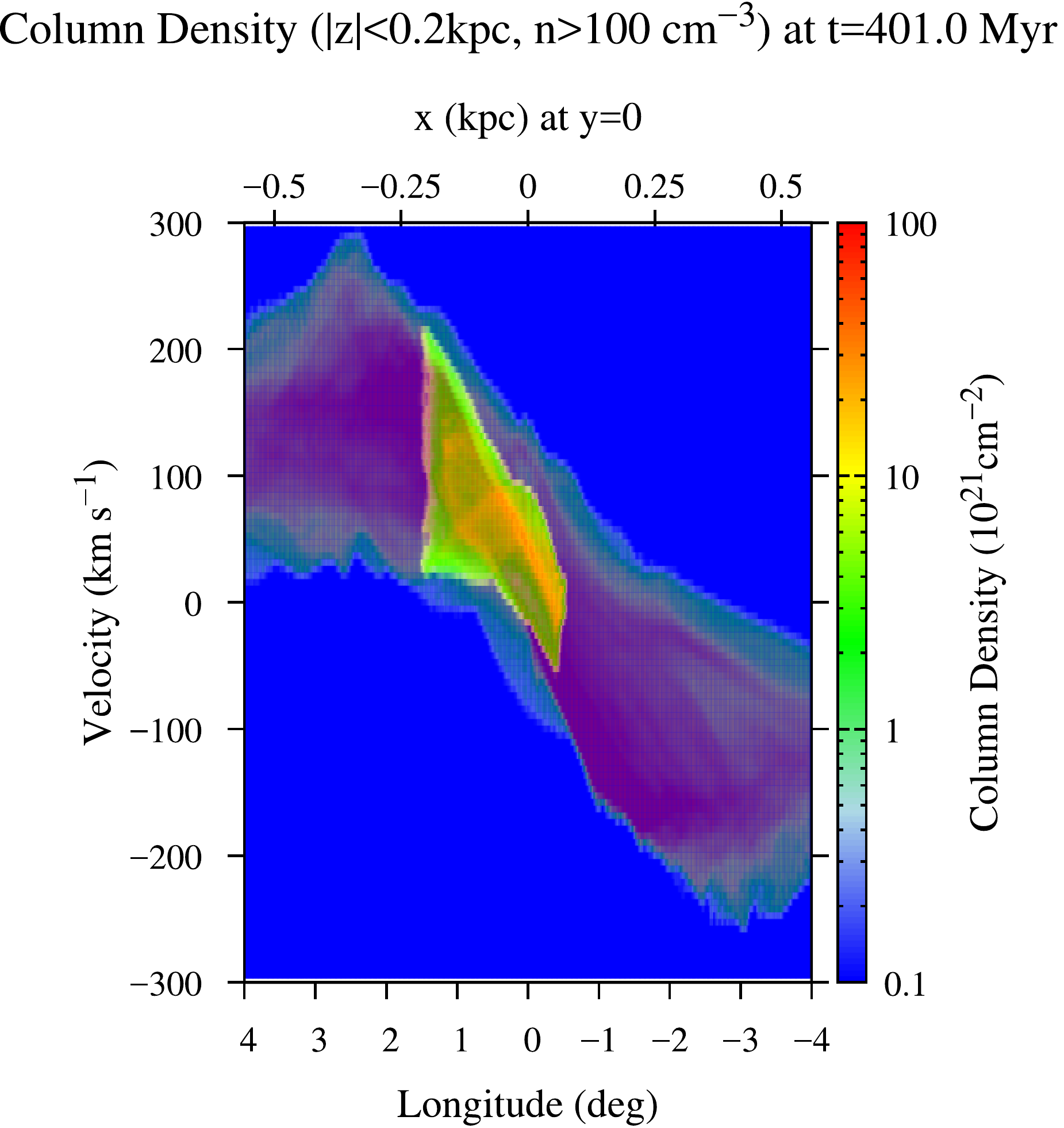}
	\caption{
		Same as the right panel of Figure \ref{fig_lv_diagram} but for different vieing angles, $\phi=250^{\circ}$ (left) and $210^{\circ}$ (right). 
	}	
	\label{fig_lv_diagram_rotation} 
\end{figure*}

Figure \ref{fig_lv_diagram_rotation} shows simulated $l-v$ diagrams observed from different angles,
$\phi=250^{\circ}$, (left), and $\phi=210^{\circ}$, (right).
These $l-v$ diagrams show that the downflow region is observed as very different features depending on the viewing angle.
For example, the right panel with $\phi=210^{\circ}$ shows the downflow region,
which is located in front of the Galactic centre, is distributed from the $+l$ region to the $-l$ region.
In this case, the velocity dispersion of the dense region
(redder in Figure \ref{fig_lv_diagram_rotation}) is smaller;
the velocity dispersion in $l-v$ diagrams depends on the viewing angle. 
Position-Velocity (P-V) diagrams are obtained for various galaxies \citep{Sakamoto2006a},
which are observed from various viewing angles.
Comparing observed P-V diagrams  of galaxies to simulated P-V diagrams,
we can understand actual configuration of flows, which is one of our future targets. 

\subsection{Non-ideal MHD effects}

\pink{
In this work we have assume the ideal MHD approximation,
which is valid if the sufficient ionization is achieved.
Otherwise, the magnetic field is not strongly coupled to the gas,
and non ideal MHD effects, Ohmic diffusion (OD), Hall effect (HE),
and ambipolar diffusion (AD), need to be taken into account.
}

\pink{
Resent observation towards the GC region showed the existence of 
a large amount of H$_3^{+}$ and H$_3$O$^{+}$  \citep{Oka2005, VanderTak2006},
which suggests an ionization rate
$\zeta\sim10^{-15}\unit{s^{-1}}$
in the GC region is much higher than the typical rate in the galactic disc
\citep[$\zeta\sim3.0\times10^{-17}\unit{s^{-1}}$, cf. ][]{McCall1998,McCall2002,Geballe1999}.
A primary source of the ionization of dense molecular gas
is cosmic rays injected from the supernovae
exploding every few or ten thousand years in the GC region
\citep{Crocker2011a,Yusef-Zadeh2013},
which can give the above ionization rate.
}

\pink{
The thermal rate coefficient of recombination of $H^{+}_3$, is estimated to be 
$\kappa=-1.3\ord{-8}+1.27\ord{-6}\ T^{-0.48} \unit{cm^3\ s^{-1}}$ \citep[cf. ][]{McCall2004}. 
	For the temperature of 150 K, $\kappa\sim1.0\ord{-7}\unit{cm^3\ s^{-1}}$.
	We can estimate the equilibrium ionization fraction $\chi_{e}=n_e/n_H$ in molecular clouds as
	\begin{align}
	\chi_e = \frac{n_e}{n_H}&= \sqrt{\frac{\zeta}{\kappa n_H}}\notag\\
	&\approx 10^{-5}
	\enc{\frac{\zeta}{10^{-15} \unit{s^{-1}}}}^{1/2}
	\enc{\frac{n_H}{100 \unit{cm^{-3}}}}^{-1/2}.
	\label{eqn_ionization}
	\end{align}
}

\pink{
In order to examine the validity of the ideal MHD approximation
in the GC region, we introduce magnetic Reynolds numbers, $Rm$, which is defined as the ratio of an inertial term to a magnetic diffusion term in the induction equation: 	
	$
	Rm\equiv{LV}/{\eta},
	$
	where $L$ and $V$ are
	typical length- and velocity- scales.
	$\eta$ is magnetic diffusion coefficient of each non-ideal MHD effects.
	In the following paragraphs, the subscript $i$ and $N$ denote ions and neutrals.
}

\pink{
	Ohmic diffusion is caused by excess electron-neutral collision, 
	which especially tends to be effective in relatively high-density regions.
	Ohmic resistivity (magnetic diffusion coefficient) is estimated \citep{Blaes1994} as
	\begin{align}
	\eta_\mathrm{OD}
	\approx 2.83\ord{11}\frac{\average{\sigma v}_{eN}}{\chi_e}\unit{cm^2\ s^{-1}}
	\approx 230 \frac{\sqrt{T_e}}{\chi_e} \unit{cm^2\ s^{-1}},\label{eqn_ohmic}
	\end{align}
	where $T_e$ is temperature of electrons,
	and the momentum transfer rate coefficient of an electron for a neutral particle is given 
	$\average{\sigma v}_\mathrm{eN}=
	8.28\ord{-10}\ T_\mathrm{e}^{1/2}\unit{cm^{3}\ s^{-1}}
	$
	\citep{Draine1983}.
	We take a typical length scale of magnetic loops in our simulation for $L$ and the sound velocity for $V$, and then
	\begin{align}
	LV=1.0 \ord{27} \enc{\frac{L}{0.33 \unit{kpc}}}\enc{\frac{V}{100\ \unit{km\ s^{-1}}}} \unit{cm^{2}\ s^{-1}}.\label{eqn_inertial}
	\end{align}
	From equations (\ref{eqn_ohmic}) and (\ref{eqn_inertial}),
	the magnetic Reynolds number of Ohmic diffusion is shown as
	\begin{align}
	Rm_\mathrm{(OD)}&=\frac{LV}{\eta_\mathrm{OD}}\notag\\
	&\approx 3.5 \ord{18}
	\enc{\frac{\chi_\mathrm{e}}{10^{-5}}}
	\enc{\frac{T_\mathrm{e}}{150 \unit{K}}}^{-1/2}\notag\\
	&\hspace{15mm}\enc{\frac{L}{0.33 \unit{kpc}}}
	\enc{\frac{V}{100\ \unit{km\ s^{-1}}}},
	\end{align}
	which is much larger than unity.
	This imply that Ohmic diffusion is negligible in the molecular gas near GC region.
}

\pink{	
	When the density is lower,
	the momentum transfer between ion species and electron species is weak.
	In this case, Hall effect,
	one of the magnetic diffusion, occurs by carrying Hall current.
	Since Hall current depends on strength of the magnetic field $B$ and electron number density $n_e$,
	Hall diffusivity is given by
	\begin{align}
	\eta_\mathrm{HE}=\frac{c}{4\pi e}\frac{B}{n_e},
	\end{align}
	where $c$ is the light speed and $e$ is the elementary charge.
	Substituting $n_e=\chi_e n$ with $n=(n_N+n_i)\sim n_N$ and $\rho=\mu m_{H}/n$ with $\mu=1.2$, we can derive
	\begin{align}
	\eta_\mathrm{HE}&\sim 5.7\ord{17}
	\enc{\frac{B}{500 \unit{\mu G}}}
	\enc{\frac{\rho}{1.0\ord{-21}}}^{-1}
	\enc{\frac{\chi_\mathrm{e}}{10^{-5}}}^{-1},
	\end{align}
	and then, the magnetic Reynold number with respect to Hall effect is shown as
	\begin{align}
	Rm_\mathrm{(HE)}&=\frac{LV}{\eta_\mathrm{HE}}\notag\\
	&\approx 1.75 \ord{9}
	\enc{\frac{B}{500 \unit{\mu G}}}^{-1}
	\enc{\frac{\rho}{1.0\ord{-21}}}
	\enc{\frac{\chi_\mathrm{e}}{10^{-5}}}\notag\\
	&\hspace{15mm}\enc{\frac{L}{0.33 \unit{kpc}}}
	\enc{\frac{V}{100\ \unit{km\ s^{-1}}}}.
	\end{align}
	Therefore Hall effect is not effective in the GC region.
}

\pink{	
	In the further lower- density region,
	because of the ineffective collision between ion and neutral particles,
	magnetic field is diffused in fluid consisting of neutral particles,
	called Ambipolar diffusion.
	Ambipolar diffusivity can be estimated as
	\begin{align}
	\eta_\mathrm{AD}=
	\frac{(m_i+m_N)B^2}{4\pi \average{\sigma v}_{iN}\rho_i\rho_N},
	\end{align}
	where the collision rate of an ion for a neutral particle is given
	$\average{\sigma v}_{iN}=1.9\ord{-9}\unit{cm^{3}\ s^{-1}}$\citep{Draine1983}.
	Assuming charge neutrality,
	\begin{align}
	{\eta_{AD}}&\approx
	2.1\ord{-16}\frac{B^2}{\rho^2 \chi_e} \unit{cm^2\ s^{-1}}\notag\\
	&= 5.24\ord{24}
	\enc{\frac{B}{500 \unit{\mu G}}}^2
	\enc{\frac{\rho}{1.0\ord{-21}}}^{-2}
	\enc{\frac{\chi_\mathrm{e}}{10^{-5}}}^{-1}\unit{cm^2\ s^{-1}}\label{eqn_amb}
	\end{align}
	From equations (\ref{eqn_inertial}) and (\ref{eqn_amb}),
	\begin{align} 
	Rm_\mathrm{(AD)}
	&=\frac{LV}{\eta_\mathrm{AD}}\notag\\
	&\approx 2.0 \ord{2}
	\enc{\frac{B}{500 \unit{\mu G}}}^{-2}
	\enc{\frac{\rho}{1.0\ord{-21}}}^{2}
	\enc{\frac{\chi_\mathrm{e}}{10^{-5}}}\notag\\
	&\hspace{15mm}\enc{\frac{L}{0.33 \unit{kpc}}}
	\enc{\frac{V}{100\ \unit{km\ s^{-1}}}},
	\end{align}
	which means that 
	the ambipolar diffusion may not be negligible
	if the ionization degree is low.
	In our future studies, we plan to include the effect of ambipolar diffusion with an appropriate estimate of ionization degree. 
}

\subsection{Turbulent Diffusion}
\begin{figure}
	\includegraphics[clip,width=\hsize]{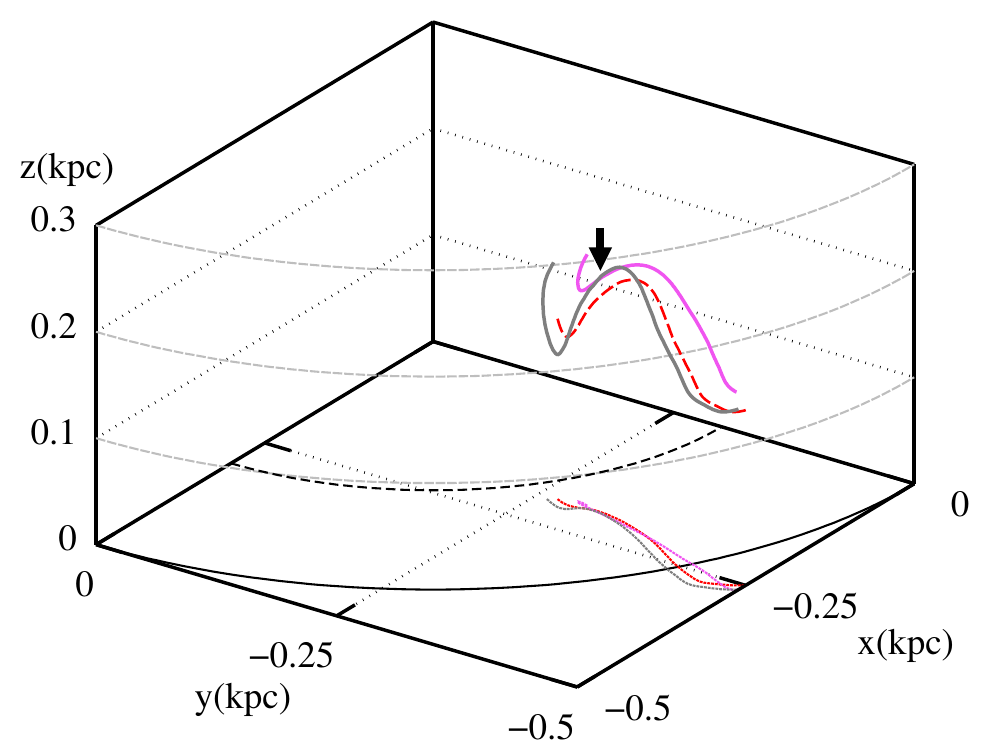}
	\caption{\pink{
			The time evolution of MF1. The red dashed line denotes the original position at $t=401.0\ \unit{Myr}$.
			The gray line is MF1 at $t=401.15\ \unit{Myr}$ transported along velocity field of the gas from the red dashed line
			The magenta line is the magnetic field line tracked from the loop top region  (arrow) of the gray line. 
			The lines on the $x-y$ plane denote the projection of each line. 
		}
	}
	\label{fig_evolve}
\end{figure}
\pink{
	Even in the ideal MHD condition,
	the motion of magnetic field could be deviated from
	the background velocity field by turbulent reconnection
	and diffusion \citep{Lazarian2009,Lazarian2015}.
	Furthermore, in MHD simulations,
	numerical diffusion could cause the decoupling
	between magnetic field and gas. 
	In Figure 13, we investigate how the numerical and turbulent diffusion affects the motion of MF1. 
	The dotted red line is MF1 at $t=401.0\ \unit{Myr}$.
	The gray line is drawn by advecting the dotted line along the velocity field with 0.15 Myr. Namely, if the MF1 is perfectly coupled with the gas, this loop should be observed as this gray line at $t=401.15$ Myr.
	The magenta line is drawn by tracking the magnetic field line from the loop top location (arrow);  this is the actual loop structure at $t=401.15$ Myr. 
	The comparison between the gray and magenta lines indicates that the footpoints of MF1 are drifting from the gas and MF1 is not so well coupled to the background gas flow because of the numerical and turbulent diffusion.
	In addition, the footpoints are drifted from the background gas flows by the transport of the angular momentum of the magnetic field.
	By these processes, the footpoints are drifted in both radial and azimuthal directions.  
	}

\section{Conclusion}

We investigated the magnetic activity in the Galactic bulge region from the 3D global MHD simulation by \citet{Suzuki2015}. 
In this paper, we particularly focused on vertical flows excited by MHD processes.  

Most notably, fast downflows are falling down along inclined magnetic fields,
which are formed as a result of rising magnetic loops by magnetic buoyancy \citep{Parker1966}.
The velocity of these downflows reaches $\approx 100$ km s$^{-1}$ near the footpoints. 
The two footpoints of rising magnetic loops are located at different radial positions.
The field lines are deformed by the differential rotation, in addition to the rising motion.
If the inner footpoint is located at a leading position with respect to the rotation at the rising phase,
the field line is stretched by the differential rotation.
On the other hand, if the inner footpoint is located at the following position at the rising phase,
a tall loop is developed because the field line is not stretched initially
by the differential rotation before the inner footpoint catches up the outer footpoint.

The angular momentum is generally transported to the outward direction
along the field lines, which
makes the radial force balance broken down. 
As a result, the gas streams inward and fall down along the
inclined field lines to the Galactic plane.
As a result, a fast downflow is often observed only near the one footpoint
located at  the inner radial position.
The gas is compressed by the downflow near the footpoint of the ``magnetic sliding slope'',
which is possibly a seed of a dense cloud \citep{Torii2010b}.

In addition, the horizontal components of the velocity are also excited
from the downward flow along the magnetic slopes,
which are observed as high velocity features in a simulated $l-v$ diagram.
On the other hand,
this simulation has not treated the thermal evolution yet, hence, in forthcoming paper,
we plan to take into account heating and cooling processes in the gas component.
This treatment enables us to study the formation of dense molecular clouds in more realistic condition.

\section*{Acknowledgement}
This work was supported in part by Grants-in-Aid for 
Scientific Research from the MEXT of Japan, 15H05694(YF) and 17H01105(TKS).




\bibliographystyle{mnras}
\bsp	
\label{lastpage}
\end{document}